\documentclass{llncs}
\usepackage{siunitx,algorithm,graphicx,amsmath, amssymb,url}
\usepackage[noend]{algpseudocode}
\usepackage[nocompress]{cite}
\usepackage[english]{babel}
\usepackage{color}
\usepackage[shortlabels]{enumitem}
\usepackage{caption}
\usepackage{xr}
\newcommand{\ignore}[1]{}
\newcommand{\vir}[1]{``#1"}

\begin{document}
	\title{On the Suitability of Neural Networks as Building Blocks for The Design of Efficient Learned Indexes \thanks{This research is funded in part by MIUR Project of National Relevance 2017WR7SHH “Multicriteria Data Structures and Algorithms: from compressed to learned indexes, and beyond”. We also acknowledge an NVIDIA Higher Education and Research Grant (donation of a Titan V GPU). %Additional support to RG has been granted by Project INdAM – GNCS Project 2020  “Algorithms, Methods and Software Tools for Knowledge Discovery in the Context of Precision Medicine”
	}}
	
	\author{Domenico Amato\inst{1} \and
		Giosu\'e Lo Bosco\inst{1} \and
		Raffaele Giancarlo\inst{1}}
	\authorrunning{F. Author et al.}
	\institute{$^1$Dipartimento di Matematica e Informatica\\ 
		Universit\'a degli Studi di Palermo, ITALY\\}
	
	\maketitle              % typeset the header of the contribution
	
	\begin{abstract}
		With the aim of obtaining time/space improvements in classic Data Structures, an emerging trend is to combine Machine Learning techniques with the ones proper of Data Structures. This new area goes under the name of \emph{Learned Data Structures}. The motivation for its study is a perceived change of paradigm in Computer Architectures that would favour the use of Graphics Processing Units and Tensor Processing Units over conventional Central Processing Units. In turn,  that would favour the use of Neural Networks as building blocks of Classic Data Structures. Indeed, Learned Bloom Filters, which are one of the main pillars of Learned Data Structures,  make extensive use of Neural Networks to improve the performance of classic Filters. However, no use of Neural Networks is reported in the realm of Learned Indexes, which is another main pillar of that new area. In this contribution, we provide the first, and much needed,  comparative experimental analysis regarding the use of Neural Networks as building blocks of Learned Indexes. The results reported here highlight the need for the design of very specialized Neural Networks tailored to Learned Indexes and it establishes solid ground for those developments. Our findings, methodologically important, are of interest to both Scientists and Engineers working in Neural Networks Design and Implementation, in view also of the importance of the application areas involved, e.g., Computer Networks and Data Bases.
	\end{abstract}

	\section{Introduction}
	Learned Data Structures is a new research area based on the combination of Machine Learning (ML) techniques with the ones proper of Data Structures, with the aim of obtaining time/space improvements in classic Data Structures.  It was initiated by \cite{kraska18case}, it has grown very rapidly \cite{Ferragina:2020book} and now it has been extended to include also Learned  Algorithms \cite{Mitz20}.     
	
	\subsection{Computer Architectures as a Motivation for Learned Data Structures}\label{sec:motiv}
	The ML models with the most potential in this area are undoubtedly Neural Networks (NNs), because of their learning power, e.g.,  \cite{Ohn19}. Unfortunately, they require prohibitive computational power. In recent years, the introduction of Graphics Processing Unit (GPU) and Tensor Processing Unit (TPU) architectures in commercial computers \cite{LawsDead,TPU}, and the deployment of highly engineered development platforms such as Tensorflow \cite{tensorflow}, has \emph{de facto} removed the computational bottleneck referred to earlier. In fact, ML Models,  and in particular NNs,  have been increasingly used in many application domains\cite{lecun2015deep}.
	
	The major strength of these new architectures is that they can parallelise maths operations made by NNs very well,  compared to a general-purpose set of instructions.  In particular, recent studies even argue that the power of the GPU can be improved by 1000x in terms of time in the next few years, while, due to Moore's Law constraints \cite{moore1965cramming}, those improvements are not seen for classic CPUs. Furthermore, a programming paradigm based on branches of the if-then-else type seems to have been overcome in favour of a paradigm that promotes straight-line mathematical operations, which can be pipelined efficiently on those modern architectures.	
	For these reasons, using ML models such as NNs, on such advanced architectures instead of the classic Data Structures, which make extensive use of branch instructions in their code, may lead to the deployment of substantially better Data Structures, with benefits in many areas such as Computer Networks and Data Bases. 
	
	Unfortunately, although the above motivation for the design of Learned Data Structures based on NNs is indeed a strong one,  the potential gain that can be achieved is either at an initial stage of assessment or it has not been assessed at all,   as we outline next.

	\subsection{From Motivation to Design and Implementation of Learned Data Structures: The role of NNs}
	\label{ssec:BFMOT}
	
	\begin{itemize}
		{
			\item{\bf Learned Bloom Filters.}  NNs have been extensively used within the design and implementation of Learned Bloom Filters, since the very start of the area of Learned Data Structure \cite{kraska18case}. W recall that, given a universe $U$ of elements,  a Bloom Filter \cite{bloom1970} is a Data Structure to solve the \emph{ Approximate Membership Problem}  for a given set $A \subset U$. That is, given a query element $x \in U$, establish whether  $x \in A$,  with a specific False Positive Rate (FPR) $\epsilon$ and zero False Negatives. Essential parameters for the evaluation of the performance of Bloom Filters are FPR, Query Time and Space Occupancy. Those parameters are very intimately connected, as well explained in \cite{broder2003bloom}. The current versions of Learned Bloom Filters that have been proposed in the literature \cite{kraska18case,Mitz18,vaidya2020partitioned,Zhenwei2020} use convolutional and recurrent neural networks. The interested reader can find details in \cite{Fumagalli:2021}, together with experimental comparative performance analysis.
			\item {\bf Learned Indexes.} Learned Indexes \cite{kraska18case} have been introduced to solve the so-called \emph{Predecessor Search Problem}. Given a sorted table $A$ of $n$ keys taken from a universe $U$, and a query element $x$ in $A$, the  Predecessor Search Problem consists in finding the  $A[j]$ such that $A[j] \leq x < A[j+1]$. A synopsis of the Learned Indexes  Methodology is presented in Section \ref{sec:models}. Although several versions of Learned Indexes have been proposed, e.g.,  \cite{amato2021learned,amato2021lncs,Ferragina:2020book,Ferragina:2020pgm,Hadian18,kraska18case,Marcus20}, surprisingly, none of them use NNs. Even more surprisingly,  no comparative performance analysis regarding the use of NNs in Learned Indexing has been carried out.
			\item{\bf Additional Learned Data Structures.} Learned Hash Functions \cite{kraska18case} as well as Learned Rank and Select Data Structures \cite{Boffa:2021} do not use NNs. Again, no performance analysis regarding those models has been carried out.
			
		}
	\end{itemize}

	\subsection{Our Results: The Role of Neural Networks in the Design of Learned Indexes - the Atomic Case}
	In view of the State of the Art reported in Section \ref{ssec:BFMOT}, our novel contribution to the advancement of the area of Learned Data Structures is the first assessment of the suitability of NNs as building blocks of Learned Indexes. Indeed, in order to provide a  clear comparative assessment of the potential usefulness of NNs in the mentioned domain, we consider Atomic Models for  Learned Indexes, defined in Section \ref{sec:AM}. Intuitively, they are the simplest models one can think of. The rationale for their choice is that, if NNs do not provide any significant advance with respect to very simple Learned Indexes, in view of the results in \cite{amato2021lncs,amato2021learned}, they have very little to offer to Learned Indexing in their current generic, if not \vir{textbook},  form. Even with the advantage of GPU processing. Technically, we offer the following contributions.
	
	%As opposed to the case of bloom filters, there is not an extensively use of NN as building blocks of the Learned Data Structures, and for the first time, we investigate on this direction giving these contributions:. 
	
	\begin{itemize}
		\item The first design of a Learned Index based on NNs Models only. We choose Feed Forward NNs because they offer a good compromise between time efficiency, space occupancy and ability to learn \cite{bishop1995neural}. Since this Learned Index has no other ML subcomponent, we refer to it as an Atomic Learned Index.
		
		\item An extensive experimental study about the effectiveness of this Atomic Learned Index, both in the case of CPU and GPU processing.

		%that uses study about the effectiveness of NNs as components of Learned Indexes both in the case of GPU and CPU processing, in particular, we access the ability of NNs to solve the Predecessor Search Problem with the aid of Binary Search routines. 
		
		%An experimental study about the effectiveness of Neural Network Architectures used as ML components of Learned Indices, both on the cases of GPU and CPU processing. 
		
		\item An extensive comparison of this Atomic Learned Index with respect to analogous Atomic Indexes that use only Linear Regression \cite{FreedmanStat} as the learned part. Those models have been formalized and studied in \cite{amato2021lncs,amato2021learned} and they are valid building blocks of more complex models used for Learned Indexes, e.g., the {\bf RMI} family \cite{Marcus20}.

		%e proposals in the previous point, with respect to elementary Learned Indexes. Indeed if their performance is no better than the one granted by simple Learned Indexes, in view of the result reported in {\color{red} lavoro IS e AIXIA}  they have very little hope to make an impact on this area.    
	\end{itemize}

	%Rivedere brevemente la letteratura sui learned indices, e poi dire "as opposed to bloom filters, non c'è nessun utilizzo dell NN come building blocks di queste strutture dati e per la prima volta noi investighiamo su questa cosa dando questi contributi."
	
	%In particular, we provide an experimental study about the effectiveness of Neural Network Architectures used as ML components of Learned Indices, in comparison with the simple Atomic ones.
	
	As opposed to Learned Bloom Filters,  our results clearly indicate the need to design NNs specifically focused on their use in the Learned Indexing paradigm. A study analogous to ours, that paves the way to developments of  NNs more specific for  Learned Indexing, is not available in the Literature and it is of Methodological importance.  
	
	The software we have used for our experiments is available at  \cite{gitnn}.
	\subsection{Organization of the Paper}
	
	This paper is organized as follows. Section \ref{sec:models} provides a synoptic description on Learned Indexing. Section \ref{sec:AM} presents the Atomic Models that we consider for this research. In Section \ref{sec:EM}, we illustrate the adopted experimental methodology. Section \ref{sec:RES} reports experiments and findings. Finally,  in Section \ref{sec:ch2Conc}, we provide conclusions and future direction of research.

	\begin{figure}[tbh]
		
		\centering
		\includegraphics[width=0.3\textwidth]{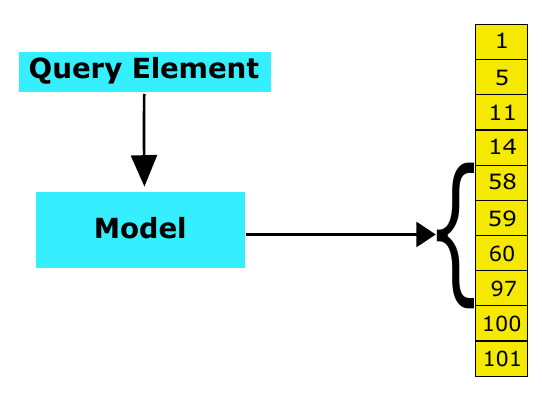}
		\caption{{\bf  A general paradigm of  Learned Searching in a Sorted Set} \cite{Marcus20}. The model is trained on the data in the table. Then, given a query element, it is used  to predict the interval in the table where to search (included in brackets in the figure).}
		%\end{center}
		\label{img:models}
	\end{figure}
	
	\section{Learned Indexes: a Synopsis}\label{sec:models}

	Consider a sorted table $A$ of $n$ keys, taken from a universe $U$.  It is well known that Sorted Table Search can be phrased as the  Predecessor Search Problem:  for a given query element $x$, return the  $A[j]$ such that $A[j] \leq x < A[j+1]$. Kraska et al. \cite{kraska18case} have proposed an approach that transforms such a problem into a learning-prediction one. With reference to Figure \ref{img:models}, the model learned from the data is used as a predictor of where a query element may be in the table.  To fix ideas, Binary Search is then performed only on the interval returned by the model. 
	
\begin{figure}[tb]
		
		%\begin{center}
		\centering
		(a)
		\begin{minipage}{0.25\textwidth}
			\includegraphics[width=\linewidth]{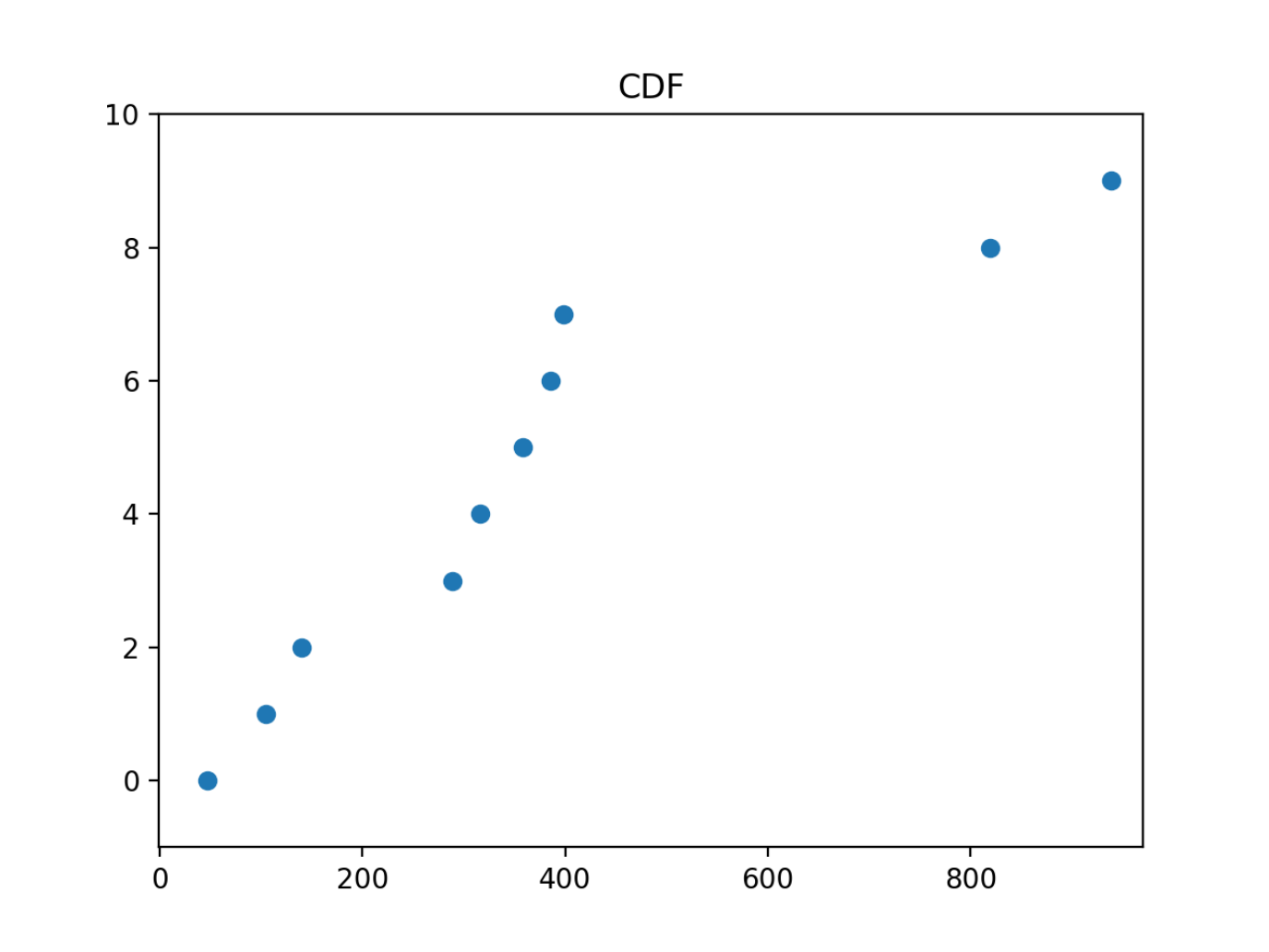}
		\end{minipage}\hfill
		(b)
		\begin{minipage}{0.25\textwidth}
			\includegraphics[width=\linewidth]{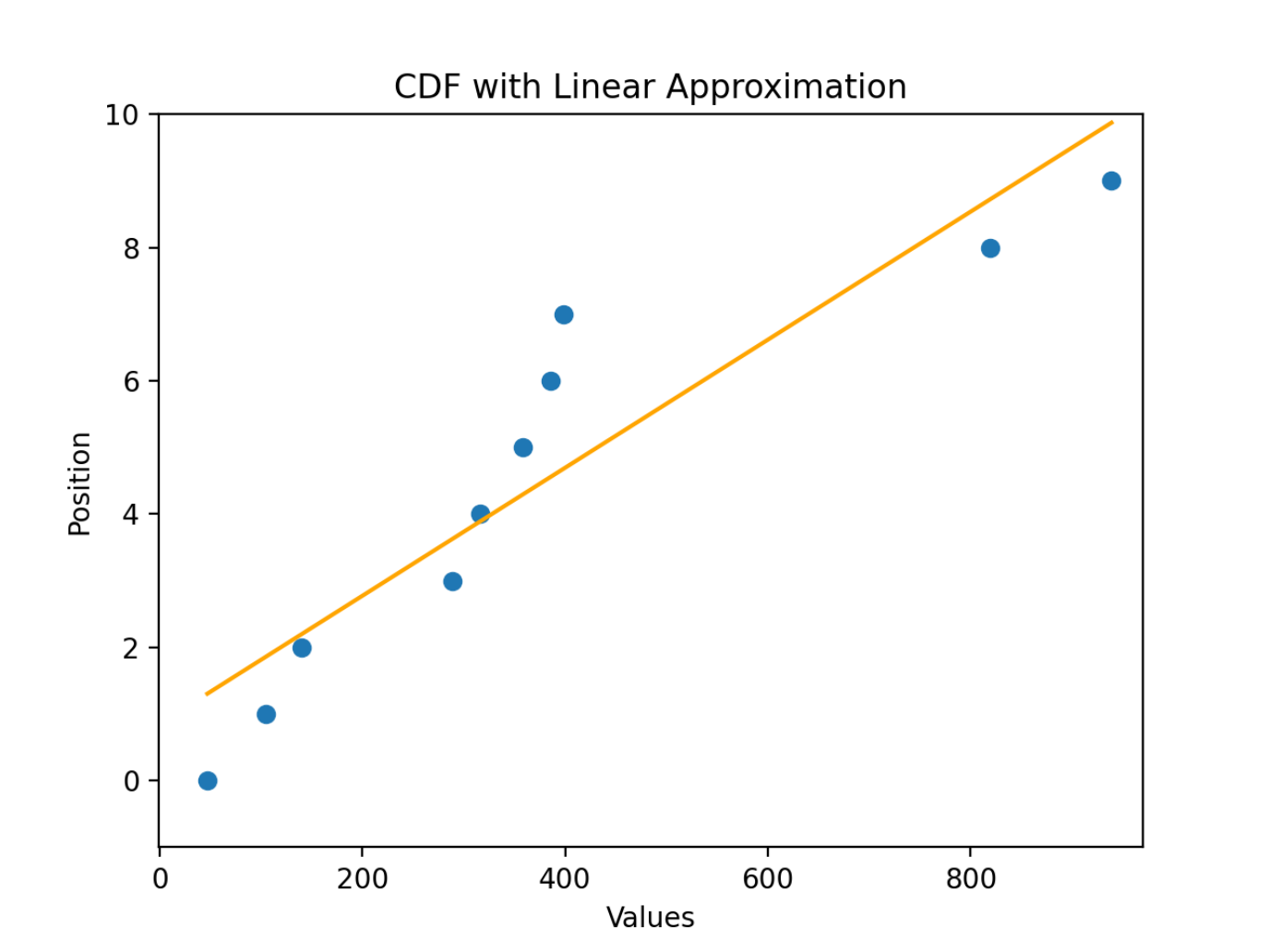}
		\end{minipage}\hfill
		(c)
		\begin{minipage}{0.25\textwidth}%
			\includegraphics[width=\linewidth]{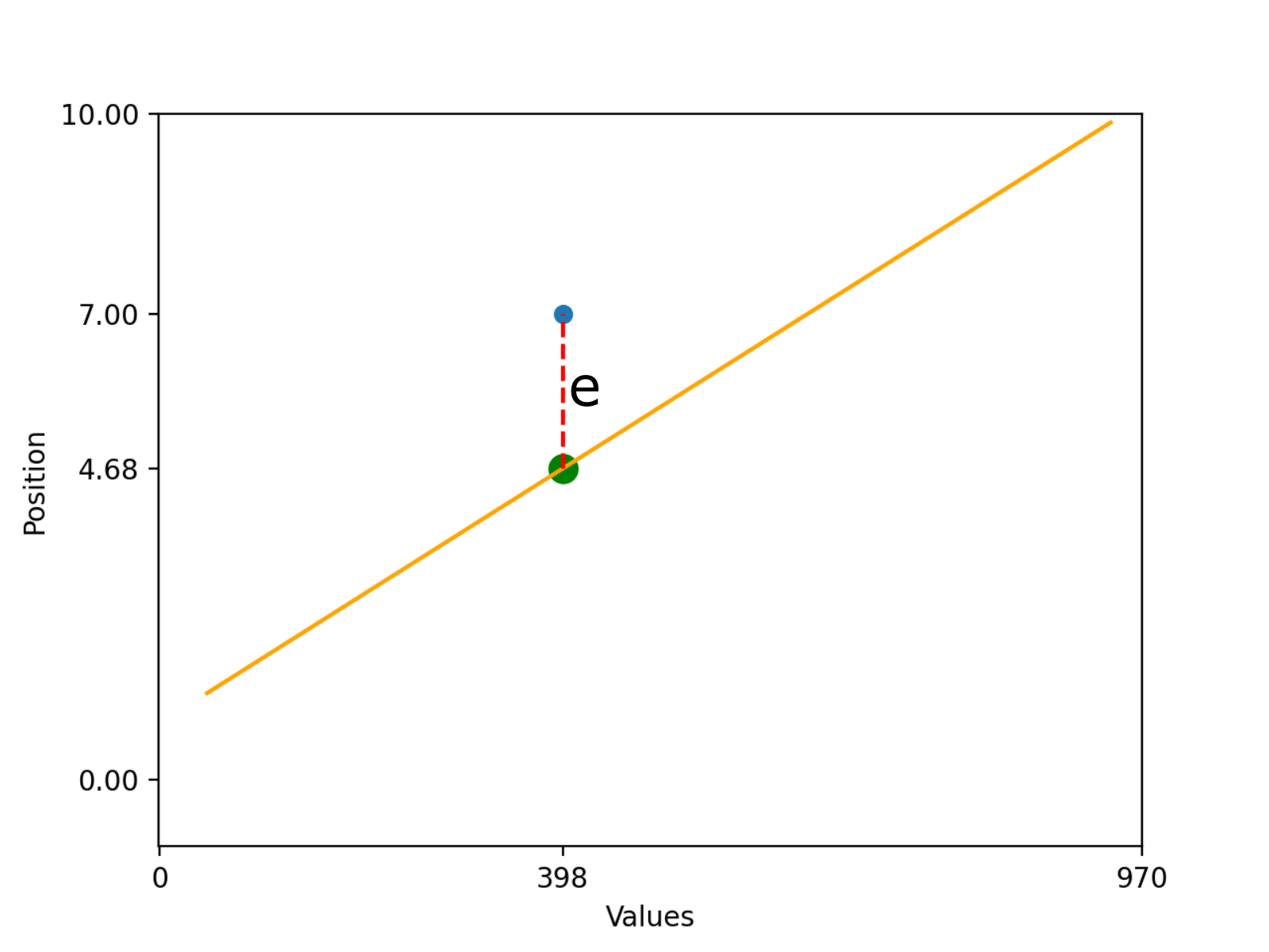}
		\end{minipage}
		\caption{{\bf  The Process of Learning a Simple Model via Linear Regression.} Let $A$ be $[47, 105, 140, 289, 316, 358, 386, 398, 819, 939]$. (a) The CDF of A. In  the diagram, the abscissa indicates the value of an element in the table, while the ordinate is its rank. (b) The  $a$ and $b$ values of the straight line  $F(x)=ax+b$ are obtained via Linear Regression, with Mean Square Error Minimization.  (c) The maximum error $\epsilon$ one can incur in using $F$ is $\epsilon=3$, i.e., accounting for rounding,  it is the maximum distance between the rank of  a point in the table and its rank as predicted by $F$. In this case, the interval to search into, for a given query element $x$,  is given by $I=[F(x)-\epsilon, F(x)+\epsilon]$. }
		\label{fig:CDF}
		%\end{center}
		
	\end{figure}

	We now outline the basic technique that one can use to build a model for $A$. It relies on Linear Regression \cite{FreedmanStat}. 
	With reference to the example in Figure \ref{fig:CDF} and assuming that one wants a linear model, i.e., $F(x)=ax+b$, Kraska et al. note that one can fit a straight line to the CDF and then use it to predict where a point $x$ may fall in terms of rank and accounting also for approximation errors. In terms of regression, the function $F$ a \emph{Model} for the CDF function. 
	
	More in general, in order to perform a query, the model is consulted and an interval in which to search for is returned. Then, to fix ideas,  Binary Search on that interval is performed.  
	Different models may use different schemes to determine the required range, as outlined in Section  \ref{sec:models}. 
	The reader interested in a rigorous presentation of those ideas can consult Markus et al. \cite{Marcus20}. 
	
	\section{Atomic Models for Learned Indexes}
	\label{sec:AM}
	As outlined in Section \ref{sec:models}, a simple linear function that provides a model for the CDF of the data is a model for Learned Indexes. With reference to Figure \ref{img:models}, when the model consists of either a closed-form formula or a simple program that provide an estimation of the CDF on a given point, we refer to it as \emph{Atomic}. That is, it has no sub-component that has been learned from the data. As stated in the Introduction, more complex models exist. However, as already outlined,  for the aim of this research,  it is best to consider Atomic Models only. In particular, we consider models that come from an analytical solution to regression problems (see Section \ref{ssec:SMR}), and models that use NNs (see Section \ref{ssec:NN}). 
	
	\subsection{CDF Function Models Based on Analytic Solutions to Regression Problems}
	
	\label{ssec:SMR}
	Regression	is a methodology for estimating a given  function $G:\mathbb{R}^m \rightarrow \mathbb{R}$  via a specific function model $\tilde{G}$. The independent variables in $x \in \mathbb{R}^m$ and the dependent variable $y \in \mathbb{R}$  are usually referred to as predictors and outcome, respectively. The parameters of $\tilde{G}$ are estimated by minimizing an error function, computed using a sample set of predictors-outcome measurements.
	The most commonly used Regression Loss Function is the Mean Square Error. Such a task can be accomplished in several ways. 
	Here we follow the method outlined in  \cite{Goodfellow-et-al-2016}. In particular, we present closed-form formulae solving the posed minimization problem, with a linear (as a matter of fact, polynomial).
	
	Linear regression ({\bf LR} for short) is the case when a geometric linear form is assumed as a model. In the case,  when $m = 1$, it is referred to as Simple Linear Regression ({\bf SLR} for short) and as Multiple Linear Regression ({\bf MLR} for short), otherwise.
	
	For the general case of {\bf LR}, given  a training set of $n$ predictor-outcome couples $(\mathbf{x}_i,y_i)$, where $\mathbf{x}_i \in \mathbb{R}^m$ and $y_i \in \mathbb{R}$, the goal is to characterize the linear function model $\tilde{G}(\mathbf{x})=\mathbf{\hat{w}} \mathbf{x}^T+\hat{b}$ by estimating the parameters $\mathbf{\hat{w}} \in \mathbb{R}^m$ and $\hat{b} \in \mathbb{R}$, using the sample set. We can define  a matrix $\mathbf{Z}$ of size $n \times (m+1)$ (usually referred to as the design matrix), where $\mathbf{Z}_i$ 
	is the $i$-th row of $\mathbf{Z}$ such that ${\mathbf{Z}_i}= [\mathbf{x}_i,1 ]$. Moreover, $\mathbf{y}$ indicates the vector of size $n$ such that the outcome  $y_j$ is its $j$-th component. The Mean Square Error  minimization on the basis of the estimation is:
	
	\begin{equation}
		{\bf MSE}(\mathbf{w},b) = {\frac{1}{n} \left \| [\mathbf{w},b]  \mathbf{Z}^T-\mathbf{y} \right \|}_2^2
		\label{eq:MSE}
	\end{equation}
	
	{\bf MSE} is a convex quadratic function on $[\mathbf{w},b]$, so that the unique values  that minimize it  can be obtained by setting its gradient $\nabla_{\mathbf{w},b}$ equal  to zero.
	The closed form solution for the parameters $\mathbf{w},b$ is  
	
	\begin{equation}
		[\mathbf{\hat{w}},\hat{b}]= \mathbf{y} \mathbf{Z} (\mathbf{Z}^T \mathbf{Z})^{-1} 
		\label{eq:PLR}
	\end{equation}
	
	It is to be noted that the  {\bf SLR} case is characterized by the choice of a polynomial of degree $g=1$. The general case of Polynomial Regression ({\bf PR} for short), using polynomials with degree $g>1$, are special cases of {\bf MLR}. Indeed, we can consider the model:
	
	$$\tilde{G}(\mathbf{z})=\sum_{i=1}^g w_i x^i + b=\mathbf{w}\mathbf{z}^T+b, $$
	
	\noindent where $w$ is of size $g$, $\mathbf{z}=[x,..,x^{g-1},x^g] \in \mathbb{R}^{g}$ is the predictor vector for {\bf  MLR}.
	
	In this paper, we use linear, quadratic and cubic regression models to approximate the function $F$ given by the $CDF$ of the data. In particular, the corresponding models  are prefixed by {\bf L, Q,} or {\bf C}. 
	
	\ignore{
		\begin{figure}[t]
			\centering
			\includegraphics[width=0.3\textwidth]{neuron.pdf}
			\caption{{\bf  A simple neuron}: The basic elements of a neural network. Here we consider the case when the scalar product is the operation and the rectified linear activation unit \emph{relu} (in red) is the activation function. The scalar product between the vector with components $x^1,..,x^d$ and the parameter vector with components $w_1,..,w_d$ is given as input to the relu. } 
			\label{img:neuron}
		\end{figure}
	}
	
	\subsection{CDF Function Models Based on Neural Networks}\label{sec:FNN}
	\label{ssec:NN}
	Another method to learn a function $G:\mathbb{R}^m \rightarrow \mathbb{R}$ is to use NNs. In particular, we focus on Feed-Forward NNs, where the general strategy consists of an iterative training phase during which an improvement of the ${\tilde{G}}$ approximation is made. Starting from an initial approximation $\tilde{G}_0$, at each step $i$,  an attempt is made to minimize an error function $E$ so that $E(\tilde{G}_{i-1}) \geq E(\tilde{G}_i)$. The minimization is carried out on a training set $T$ of examples.   	 
	The process can stop after a fixed number of steps or when, given a tolerance $\delta$, $\mid E(\tilde{G}_{i-1}) - E(\tilde{G}_i) \mid \leq \delta$.
	In the following,  we report the basic elements that characterize the type of  NN we use.
	
	\begin{enumerate}
		\item {\bf ARCHITECTURE TOPOLOGY}.
		\begin{enumerate}
			\item As atomic element of our   NN, we use a standard  Perceptron \cite{bishop1995neural}. with  \emph{relu} activation function.
			\item The number of Hidden Layers $H$.
			\item For each hidden layer $h_i$, its  number of Perceptrons $n_{h_i}$.
			\item The connection between each layer. In our case, a Fully Connected NN is used, i.e. each Perceptron  of layer $h_i$ is connected with each Perceptron  of the next layer $h_{i+1}$.
		\end{enumerate}
		
		\item {\bf THE LEARNING ALGORITHM}.
		\begin{enumerate}
			\item The error function $E$, that is used to measure how close is $\tilde{G}$ to $G$.
			\item The gradient descent iterative process that starts from a $\tilde{G}_0$ and, at each step, better approximates $G$ reducing $E$, changing layer by layer, by a backward and forward pass, the parameters of each layer. It can be characterized by a learning rate, i.e. the multiplicative constant of the gradient error. 
		\end{enumerate}
		
		\item {\bf THE TRAINING SCHEME}.
		\begin{enumerate}
			\item The size of a batch $B$, i.e.,  the number of elements to extract from the training set $T$. At each extraction of the batch, the parameters are updated.
			\item The number of epochs $ne$ that corresponds to the number of times the training set $T$ is presented to the NN for the minimization of $E$. 	   	  
		\end{enumerate}

	\end{enumerate}
	
	The learning algorithm of a NN uses a proper training set to perform the gradient descent iterative process. For the purpose of indexing, which is our goal, the training data are in the form of scalar integers. To perform a Regression using a NN, it is mandatory to represent the scalar integer $x$ with a vector representation $\overrightarrow{x}$. In this work,  $\overrightarrow{x}$ is a string containing the 64-bit binary representation of $x$, as suggested also by Kraska et al. \cite{kraska18case}.
	
	\ignore{The regression models basically are able to reduce the search interval where to apply a standard binary search. In addition to this Search method, in this work, we use the best ones that come out of the work by Khuong and Morin \cite{Morin17} and by Shutz et al.  \cite{Shutz18}.
		As for terminology, we follow the one in \cite{Morin17}. Indeed, we refer to standard Binary Search as  Branchy Binary Search ({\bf BBS}, for short). Moreover, we refer to Uniform Binary Search \cite{KnuthS} and its homologous routines as branch-free {\bf BFS}. Those routines differentiate themselves from the Standard one because there is no test for exit within the main loop.
		In the following, we describe two approaches to the possibility of estimating $F$, which we indicate as \emph{atomic models}.

		A learned index can assume a more complex form than the one characterized by the atomic models (Fig. \ref{img:models}(a)). This is the case of the so-called Recursive Model Index (denoted with {\bf RMI}). RMI is a multi-level, directed graph, with Atomic Models at its nodes. When searching for a given key and starting with the first level, a prediction at each level identifies the model of the next level to use for the next prediction. This process continues until a final level model is reached. This latter is used to predict the interval to search into. An  example is  provided in Fig. \ref{img:models}(b). It is to be noted that Atomic Models are {\bf RMI}s.  }

	\subsection{Prediction Accuracy of an Atomic Model}\label{sec:RF}
	
	As well illustrated in Figure \ref{fig:CDF}, the approximation error is very important in reducing the size of the interval to be searched into. Smaller the error, the smaller the portion of the table in which the final search must be performed. In this paper, we characterize the accuracy in prediction of a model via the \emph{reduction factor}  ({\bf RF}): the percentage of the table that is no longer considered for searching after a prediction. Because of the diversity across models to determine the search interval, and in order to place all models on a par, we estimate empirically the {\bf RF} of a model. That is, with the use of the model and over a batch of queries, we determine the length of the interval to search into for each query (the interval $I$ in Figure \ref{fig:CDF}). Based on it, it is immediate to compute the reduction factor for that query. Then, we take the average of those reduction factors over the entire set of queries as the reduction factor of the model for the given table.

	\section{Experimental Methodology}\label{sec:EM}
	
	\subsection{Hardware and Datasets}
	\label{sec:datasets}
	Experiments have been performed using a workstation equipped with an Intel Core i7-8700 3.2GHz CPU and an Nvidia Titan V GPU. The total amount of system memory is 32 Gbyte of DDR4. The GPU is also supplied with its own 12 Gbyte of DDR5 memory and adopts a CUDA parallel computing platform. CPU and GPU are connected with a PCIe 3 bus with a bandwidth of 32Gbyte/s. The operating system is Ubuntu LTS 20.04.

	We have used both synthetic and real datasets, taken from previous studies on Learned Indexes \cite{kraska18case, Ferragina:2020pgm}. The  synthetic ones  have been generated using random sampling in $[1, 2^{r-1}-1]$,  with $r=64$.  Datasets are sorted and without duplicates. We anticipate that, as evident from the analysis in Section \ref{sec:RES}, the use of GPU training for the NNs severely limits the size of the datasets that we can use.

	\begin{enumerate}
		\item {\bf Uni} that contains data sample from a Uniform distribution defined as  
		\begin{equation}
			U(x,a,b) = \Bigg\{ 
			\begin{array}{ll}
				\frac{1}{b-a}  & \mbox{if } x \in [a,b] \\
				0 & otherwise
			\end{array}
		\end{equation} 
		where $a= 1$ e $b = 2^{r-1} -1$. Its size is  1.10e+04 Kb and it contains 1.05e+06  integers.
		
		\item {\bf Logn} that contains data sample from a Log-normal distribution defined as 
		\begin{equation}
			L(x, \mu, \sigma) = \frac{e^{-\frac{(ln x - \mu)^2}{2\sigma^2}}}{x\sqrt{2}\pi\sigma}
		\end{equation}	
		where $\mu = 0$ e $\sigma = 1$ are respectively mean and variance of the distribution. Its size is  1.05e+04 Kb and it contains 1.05e+06  integers.
	\end{enumerate}

	\begin{enumerate}	
		\item {\bf Real-wl} that contains timestamps of about 715M requests performed by a web server during 2016. Its size is 3.48e+05 Kb and it contains 3.16e+07 integers.
		\item {\bf Real-iot} that consists of timestamps of about 26M events recorded during 2017 by IoT sensors deployed in academic buildings. Its size is 1.67e+05 Kb and it contains 1.52e+07 integers.
		
	\end{enumerate}
	As for the query dataset, for each of the above tables, it has a size equal to 50\% of the reference table and contains, in equal parts, both elements present and not present in the table. For all the experiments, the query datasets are not sorted.
	\ignore{
		\begin{table}[tbh]
			\centering
			\scriptsize
			\begin{tabular}{|l|l|l|l|}
				\hline
				\multicolumn{4}{|c|}{\textbf{Uniform Distribution}}          \\ \hline \hline
				\textbf{Name}   & \textbf{Size (KB)} & \textbf{Items} & \textbf{Type} \\ \hline
				{\bf uni }    &    1.10e+04       &  1.05e+06   & integer              \\ \hline \hline
				\multicolumn{4}{|c|}{\textbf{Log-normal Distribution}}        \\ \hline \hline
				\textbf{Name}   & \textbf{Size (KB)} & \textbf{Items} & \textbf{Type} \\ \hline 
				{\bf logn }     &  1.05e+04  &  1.05e+06  &integer\\ \hline \hline
				\multicolumn{4}{|c|}{\textbf{Real Distribution}}             \\ \hline \hline
				\textbf{Name}   & \textbf{Size (KB)} & \textbf{Items} & \textbf{Type} \\ \hline
				{\bf real-wl}         &3.48e+05&3.16e+07& integer\\ \hline
				%real-itlong &  \num{1.61E+06}  &\num{1.66E+08}&float\\ \hline
				{\bf real-iot }             &  1.67e+05 &1.52e+07&integer\\ \hline
			\end{tabular}
			\caption{{\bf A summary of the Datasets}.  For each datasets in the collection,  it is shown: the name used (column \textbf{Name}), its size in Kilobyte (column \textbf{Size (KB)}), the number of elements  in it (column  \textbf{Items}), and the type of its elements (final column \textbf{Type}).\label{tab:Chp2Dataset}}
		\end{table}
	}
	\subsection{Binary Search and the Corresponding Atomic Learned Indexes}\label{sec:methods}
	For the final search stage, in addition to a standard Binary Search method, we use also Uniform Binary Search \cite{KnuthS} (see also \cite{Morin17,Shutz18}). Indeed, based on work by Khuong and Morin \cite{Morin17}, it can be streamlined to avoid \vir{branchy} instructions in its implementation. Such a streamlining results in a speed-up in regard to the standard procedure. The interested reader can find details in \cite{Morin17}. 
	We refer to the standard procedure as  Branchy Binary Search ({\bf BBS}) and to the other version as a Branch-Free ({\bf BFS}). 
	
	As for Atomic Models, we use {\bf L, Q,} and  {\bf C}. Additional Atomic Models are obtained via NNs. Indeed, we consider three types of NNs with different hidden layers, as specified next. {\bf NN0} for zero hidden layers, {\bf NN1} for one hidden layer, {\bf NN2} for two hidden layers, each layer consisting in 256 units.  
	
	Each of the above models provides two Atomic Learned Indexes, one for each Binary Search routine used for the final search stage. 
	
	\ignore{
		
		Because of the diversity across models to determine the search interval, and in order to place all models on a par, we estimate the reduction factor ({\bf RF}) of a model, i.e., the percentage of the table that is no longer considered for searching after a prediction, empirically. That is, with the use of the model and over a batch of queries, we determine the length of the interval to search into for each query (the interval $I$ mentioned at the beginning of Section \ref{sec:models}). Based on it, it is immediate to compute the reduction factor for that query. Then, we take the average of those reduction factors over the entire set of queries as the reduction factor of the model for the given table.}
	
	\section{Experiments and Findings}\label{sec:RES}
	
	We use the datasets described in Section \ref{sec:datasets}. Both training and query datasets are transformed,  as outlined in Section \ref{sec:FNN}, to use them as input of the NNs. 
	For NN training,  we use the highly-engineered Tensorflow platform,  with GPU  support. The results are reported in Section \ref{ssec:RES-T}. 
	
	As for queries, we perform the following experiments.
	
	\begin{itemize}
		\item {\bf TensorFlow}. We have carried out query experiments by uploading the platform to GPU,  in order to perform query searches with the Learned Indexes based on NNs. Because of the overhead to upload Tensorflow in the GPU, results are very disappointing and therefore not reported. This is in agreement with considerations in \cite{kraska18case}.
		
		\item {\bf NVIDIA CUDA Implementation}. We use our own implementation of the Learned Index corresponding to {\bf NN0} and with {\bf BBS} for the final search stage. The results are reported in Section \ref{ssec:QGC}. As discussed in that Section, this experiment indicates that the use of the GPU is not advantageous, even with respect to the baseline {\bf BBS}, implemented in a parallel version. Therefore, no further experiment on the GPU has been performed. 
		
		\item {\bf CPU}. In this case, we have performed the entire set of experiments. For conciseness, we report the results only with the use of {\bf BFS}, since the ones involving {\bf BBS} would add very little to the discussion. They are reported and discussed in Section \ref{ssec:RES-Q}.

	\end{itemize}

	\subsection{Training:  GPU vs CPU}\label{ssec:RES-T}
	
	%RF cntrolalre se viene dato
	%Dire che il Traing times è per element
	In Table \ref{tab:Train}, we report the training times per element for each method described in the preceding Section, and we also indicate the respective {\bf RF}, computed as indicated in Section \ref{sec:RF}.
	For what concerns Atomic Models {\bf L}, {\bf Q} and {\bf C},  the training time is the time needed to solve Eq. \ref{eq:PLR}. Regarding the NN models, the used learning algorithm is stochastic gradient descent with momentum parameter equal to $0.9$ and learning rate equal to $0.1$. The Batch size is 64, and the number of epochs is 2000.
	
	As is evident from the results reported in Table \ref{tab:Train}, even with GPU support and the use of the highly-engineered Tensorflow platform,  NNs are not competitive with respect to the {\bf L},  {\bf Q} and {\bf C}  Atomic Models, both in training time and {\bf RF}. Indeed, for each dataset, the NNs training time per item is four orders of magnitude higher than the one obtained with the non-NN Atomic Models, with comparable {\bf RF}.
	
	\begin{table}[t]
		\centering
		\scriptsize
		\begin{tabular}{|c|c|c|c|c|c|c|c|c|}
			\hline
			& \multicolumn{2}{c|}{{\bf uni}} & \multicolumn{2}{c|}{{\bf logn}} & \multicolumn{2}{c|}{{\bf real-wl} }& \multicolumn{2}{c|}{{\bf real-iot}} \\ \hline
			& {\bf TT (s)} & {\bf RF (\%)} & {\bf TT (s)} & {\bf RF (\%)} & {\bf TT (s)} & {\bf RF (\%)} & {\bf TT (s)} & {\bf RF (\%)}\\ \hline
			{\bf NN0} & 2.55e-04 & 94.08\% & 1.39e-04 & 54.40\% & 2.50e-04 & 99.99\% & 1.28e-04 & 89.90\% \\ \hline
			{\bf NN1} & 4.18e-04 & 99.89\% & 3.79e-04 & 94.21\% & 2.31e-04 & 99.88\% & 4.20e-04 & 98.54\% \\ \hline
			{\bf NN2} & 4.49e-04 & 99.87\% & 8.60e-04 & 97.14\% & 2.33e-04 & 99.8\% & 3.57e-04 & 97.31\% \\ \hline
			{\bf L} & 8.20e-08 & 99.94\% & 5.61e-08 & 77.10\% & 5.82e-08 & 99.99\% & 7.70e-08 & 96.48\% \\ \hline
			{\bf Q} & 1.27e-07 & 99.98\% & 1.02e-07 & 90.69\% & 1.14e-07 & 99.99\% & 1.25e-07 & 99.1\% \\ \hline
			{\bf C} & 1.84e-07 & 99.97\% & 1.74e-07 & 95.76\% & 1.24e-07 & 99.45\% & 1.63e-07 & 98.87\% \\\hline
		\end{tabular}
		
		\caption{ {\bf Atomic Indexes Training Time and Reduction Factor}. For each datasets and each model, it is shown: the training time per element expressed in seconds (column {\bf TT (s)}) and the percentage of the table reduction (column {\bf RF (\%)}), as described in Section \ref{sec:RF}.}\label{tab:Train}
	\end{table}

\subsection{Query: GPU Only  for NNs}\label{ssec:QGC}

We perform an experiment, only on {\bf NN0} and {\bf uni}, to see if there could be a real advantage from using the GPU for queries. In Table \ref{tab:QGC}, we report the query time per element resulting from this experiment. 
As evident from that Table, on GPUs, the copy operations from CPU to GPU, and vice versa, cancel the one order of magnitude speed-up of the maths operations. In addition, a classic parallel Binary Search {\bf BBS} on the GPU is by itself faster than its Learned counterparts, making the use of NNs on this architecture unnecessary.

%From these results, we conclude that the use of a GPU architecture for Learned Indexes is still premature, due to a data transfer bottleneck. We point out that the identification of this problem is new in the Literature.

\begin{table}[t]
	\centering
	\scriptsize
	\begin{tabular}{|c|c|c|c|c|}
		\hline
		{\bf Methods} & {\bf Copy (s)} & {\bf Op. (s)} & {\bf Search (s)} & {\bf Query (s)} \\ \hline
		{\bf NN0-BBS}  & 3.27e-08  & 4.20e-09 & 1.84e-09 & 3.27e-08 \\ \hline
		{\bf BBS }  & 2.55e-09	 & - & 1.89e-09 & 4.44e-09  \\ \hline
	\end{tabular}
	\caption{{\bf Query Time on GPUs}. {\bf NN0-BBS} refers to Binary Search with {\bf NN0} as the prediction step, while {\bf BBS} is the parallel Binary Search executed on GPU without a previous prediction. For each of these methods executed on GPU, we report: the time for CPU-GPU, and vice versa, copy operations (column {\bf Copy (s)}), the time for maths operation (column {\bf Op. (s)}), the time to search into the interval (column {\bf Search (s)}) and the total time to complete the query process (column {\bf Query (s)}). Every time in the Table is per element and is expressed in seconds.}\label{tab:QGC}
	%The first column indicates the method used to perform the search in the sorted set. The second report the time for the CPU-GPU and vice versa copy, while the third column the time used to perform {\bf NN}s math operations. The fourth report the time taken by the Binary Search, while the fifth is the total time to retrieve the element. Every time in the table is per element and is expressed in seconds.}\label{tab:QGC}
\end{table}

%Tabella della TPU poi (1) tempo veloce ma I/O Bound
%Tensorflow non va bene per le query

\subsection{Query: CPU Only for All Atomic Models}\label{ssec:RES-Q}

The query experiments results are summarized in Tables \ref{tab:NNP} and \ref{tab:RP}. As we can see, NNs are also not competitive for the query phase.

The query time on {\bf NN1} and {\bf NN2} is even two orders of magnitude greater than the one obtained with the simple {\bf L} Model. In addition, in some cases, the transformed dataset is too large to be stored entirely in the CPU memory, causing a space allocation error.

\begin{table}[t]
\centering
\scriptsize
\begin{tabular}{|c|l|l|l|l|}
	\hline
	\multicolumn{1}{|c|}{\textbf{Dataset}}  &
	\multicolumn{1}{|c|}{{\bf BFS}} &
	\multicolumn{1}{|c|}{\textbf{NN0-BFS}}  &
	\multicolumn{1}{c|}{\textbf{NN1}} & \multicolumn{1}{c|}{\textbf{NN2}} \\\hline
	{\bf uni }  & 2.81e-07 & 1.31e-07 & 1.56e-06 & 5.16e-06  \\ \hline
	{\bf logn }  & 2.08e-07 & 1.92e-07 & 1.69e-06 & 5.24e-06   \\ \hline
	{\bf real-wl }  & 3.38e-07 & 4.59e-07 & Space Error & Space Error   \\ \hline
	{\bf real-iot }  & 3.07e-07 & 4.76e-07 & 1.90e-06 & 1.94e-05   \\ \hline
\end{tabular}
\caption{{\bf CPU Prediction Effectiveness-NN Atomic Models}. \textbf{NN0-BFS}  refers to Binary Search with 
	{\bf NN0} as the prediction step, while the other two columns refer to the time taken by {\bf NN1}  and {\bf NN2} to predict the search interval only. The time is reported as time per query in second. When the model and the queries are too big to fit in the main memory, a space error is reported.}\label{tab:NNP}
\end{table}

\begin{table}[t]
\centering
\scriptsize
\begin{tabular}{|c|l|l|l|l|}
	\hline
	\multicolumn{1}{|c|}{\textbf{Dataset}}  &
	\multicolumn{1}{c|}{\textbf{BFS}} &
	\multicolumn{1}{c|}{\textbf{L-BFS}} &
	\multicolumn{1}{c|}{\textbf{Q-BFS}} &
	\multicolumn{1}{c|}{\textbf{C-BFS}} \\ \hline
	{\bf uni } & 2.81e-07 & 9.42e-08  & 8.11e-08 & 9.39e-08 \\\hline
	{\bf logn } & 2.08e-07 & 1.60e-07 & 1.59e-07 & 1.54e-07 \\\hline
	{\bf real-wl } & 3.38e-07 & 5e05e-08 & 2.12e-7 & 1.80e-7 \\\hline
	{\bf real-iot } & 3.07e-07 & 8.32e-08 & 1.99e-7 & 2.57e-7 \\\hline
\end{tabular}
\caption{{\bf CPU Prediction Effectiveness-Non NN Atomic Models}. The Table reports results with Linear, Quadratic and Cubic models. The Legend is as in Table  \ref{tab:NNP}. }\label{tab:RP}
\end{table}

\section{Conclusions}\label{sec:ch2Conc}
A perceived paradigm shift is one of the motivations for the introduction of the Learned Indexes. Despite that, the use of a GPU architecture for Learned Indexes based on NNs seems not to be appropriate when we use generic NNs as we have done here. It is to be pointed out that certainly, the use of a GPU accelerates the performance of maths operations, but the data transfer between CPU and GPU is a bottleneck in the case of NNs: not only data but also the size of the model matters.  When we consider CPU only, NN models are not competitive with very simple models based on Linear Regression. 

This research clearly points to the need to design  NN architectures specialized for Learned Indexing, as opposed to what happens for Bloom Filters where generic NN models guarantee good performance to their Learned versions. In particular, those new NN models must be competitive with the Atomic Models based on Linear Regression, which are widely used as building blocks of more complex models \cite{amato2021lncs,amato2021learned,Marcus20}. This study provides solid grounds and valuable indications for the future development of Learned Data Structures, which would include a pervasive presence of NNs.

\bibliographystyle{plain}
\bibliography{references}

\begin{thebibliography}{10}

\bibitem{gitnn}
\url{https://github.com/DomenicoAmato01/A-Benchmarking-Platform-for-Atomic-Learned-Indexes}.

\bibitem{tensorflow}
M.~Abadi.
\newblock Tensor{F}low: {L}arge-scale {M}achine {L}earning on {H}eterogeneous
  {D}istributed {S}ystems.
\newblock \url{http://download.tensorflow.org/paper/whitepaper2015.pdf}, 2015.

\bibitem{amato2021learned}
D.~Amato, G.~Lo Bosco, and R.~Giancarlo.
\newblock Learned {S}orted {T}able {S}earch and {S}tatic {I}ndexes in {S}mall
  {M}odel {S}pace.
\newblock {\em CoRR}, abs/2107.09480, 2021.

\bibitem{amato2021lncs}
D.~Amato, G.~Lo Bosco, and R.~Giancarlo.
\newblock Learned {S}orted {T}able {S}earch and {S}tatic {I}ndexes in {S}mall
  {M}odel {S}pace ({E}xtended {A}bstract).
\newblock In {\em Proc. of the 20-th Italian Conference in Artificial
  Intelligence (AIxIA), to appear in Lecture Notes in Computer Science}, 2021.

\bibitem{bishop1995neural}
C.M. Bishop.
\newblock {\em {Neural {N}etworks for {P}attern {R}ecognition}}.
\newblock Oxford University Press, USA, 1995.

\bibitem{bloom1970}
B.~H. Bloom.
\newblock Space/{T}ime {T}rade-offs in {H}ash {C}oding with {A}llowable
  {E}rrors.
\newblock {\em Commun. ACM}, 13(7):422–426, 1970.

\bibitem{Boffa:2021}
A.~Boffa, P.~Ferragina, and G.~Vinciguerra.
\newblock A ``{L}earned'' {A}pproach to {Q}uicken and {C}ompress
  {R}ank/{S}elect {D}ictionaries.
\newblock In {\em Proceedings of the SIAM Symposium on Algorithm Engineering
  and Experiments (ALENEX)}, 2021.

\bibitem{broder2003bloom}
A.~Broder and M.~Mitzenmacher.
\newblock {Network {A}pplications of {B}loom {F}ilters: A {S}urvey}.
\newblock {\em Internet Mathematics}, 1(4):485--509, 2003.

\bibitem{Zhenwei2020}
Z.~Dai and A.~Shrivastava.
\newblock Adaptive {L}earned {B}loom {F}ilter ({Ada-BF}): {E}fficient
  {U}tilization of the {C}lassifier with {A}pplication to {R}eal-{T}ime
  {I}nformation {F}iltering on the {W}eb.
\newblock In H.~Larochelle, M.~Ranzato, R.~Hadsell, M.~F. Balcan, and H.~Lin,
  editors, {\em Advances in {N}eural {I}nformation {P}rocessing {S}ystems},
  volume~33, pages 11700--11710. Curran Associates, Inc., 2020.

\bibitem{Ferragina:2020book}
P.~Ferragina and G.~Vinciguerra.
\newblock Learned {D}ata {S}tructures.
\newblock In {\em Recent {T}rends in {L}earning {F}rom {D}ata}, pages 5--41.
  Springer International Publishing, 2020.

\bibitem{Ferragina:2020pgm}
P.~Ferragina and G.~Vinciguerra.
\newblock The {PGM-{I}ndex}: {A} {F}ully-{D}ynamic {C}ompressed {L}earned
  {I}ndex with {P}rovable {W}orst-case {B}ounds.
\newblock {\em {PVLDB}}, 13(8):1162--1175, 2020.

\bibitem{FreedmanStat}
D.~Freedman.
\newblock {\em Statistical {M}odels: {T}heory and {P}ractice}.
\newblock Cambridge University Press, 2009.

\bibitem{Fumagalli:2021}
G.~Fumagalli, D.~Raimondi, R.~Giancarlo, D.~Malchiodi, and M.~Frasca.
\newblock On the {C}hoice of {G}eneral {P}urpose {C}lassifiers in {L}earned
  {B}loom {F}ilters: {A}n {I}nitial {A}nalysis within {B}asic {F}ilters.
\newblock In {\em Proceedings of the 11th International Conference on Pattern
  Recognition Applications and Methods (ICPRAM)}, pages 675--682, 2022.

\bibitem{Goodfellow-et-al-2016}
I.~Goodfellow, Y.~Bengio, and A.~Courville.
\newblock {\em Deep {L}earning}.
\newblock The MIT Press, 2016.

\bibitem{Hadian18}
A.~Hadian and T.~Heinis.
\newblock Interpolation-friendly {B}-trees: {B}ridging the {G}ap {B}etween
  {A}lgorithmic and {L}earned {I}ndexes.
\newblock In {\em Advances in Database Technology - 22nd International
  Conference on Extending Database Technology, {EDBT} 2019, Lisbon, Portugal,
  March 26-29, 2019}, pages 710--713, 2019.

\bibitem{Morin17}
P.V. Khuong and P.~Morin.
\newblock Array {L}ayouts for {C}omparison-{B}ased {S}earching.
\newblock {\em J. Exp. Algorithmics}, 22:1.3:1--1.3:39, 2017.

\bibitem{KnuthS}
D.~E. Knuth.
\newblock {\em The {A}rt of {C}omputer {P}rogramming, Vol. 3 ({S}orting and
  {S}earching)}, volume~3.
\newblock Addison-Wesley, 1973.

\bibitem{kraska18case}
T.~Kraska, A.~Beutel, E.~H Chi, J.~Dean, and N.~Polyzotis.
\newblock The {C}ase for {L}earned {I}ndex {S}tructures.
\newblock In {\em Proceedings of the 2018 International Conference on
  Management of Data}, pages 489--504. ACM, 2018.

\bibitem{lecun2015deep}
Y.~LeCun, Y.~Bengio, and G.~Hinton.
\newblock Deep {L}earning.
\newblock {\em Nature}, 521(7553):436, 2015.

\bibitem{Marcus20}
R.~Marcus, A.~Kipf, A.~van Renen, M.~Stoian, S.~Misra, A.~Kemper, T.~Neumann,
  and T.~Kraska.
\newblock Benchmarking {L}earned {I}ndexes.
\newblock {\em Proc. VLDB Endow.}, 14(1):1–13, sep 2020.

\bibitem{Mitz18}
M.~Mitzenmacher.
\newblock A {M}odel for {L}earned {B}loom {F}ilters and {O}ptimizing by
  {S}andwiching.
\newblock In S.~Bengio, H.~Wallach, H.~Larochelle, K.~Grauman, N.~Cesa-Bianchi,
  and R.~Garnett, editors, {\em Advances in Neural Information Processing
  Systems}, volume~31. Curran Associates, Inc., 2018.

\bibitem{Mitz20}
M.~Mitzenmacher and S.~Vassilvitskii.
\newblock {A}lgorithms with {P}redictions.
\newblock In TimEditor Roughgarden, editor, {\em Beyond the Worst-Case Analysis
  of Algorithms}, page 646–662. Cambridge University Press, 2021.

\bibitem{moore1965cramming}
G.~E. Moore.
\newblock Cramming {M}ore {C}omponents {O}nto {I}ntegrated {C}ircuits.
\newblock {\em Electronics}, 38(8), April 1965.

\bibitem{Ohn19}
I.~Ohn and Y.~Kim.
\newblock Smooth {F}unction {A}pproximation by {D}eep {N}eural {N}etworks with
  {G}eneral {A}ctivation {F}unctions.
\newblock {\em Entropy}, 21(7):627, 2019.

\bibitem{TPU}
K.~Sato, C.~Young, and D.~Patterson.
\newblock An {I}n-{D}epth {L}ook at {G}oogle’s {F}irst {T}ensor {P}rocessing
  {U}nit ({TPU}).
\newblock
  \url{https://cloud.google.com/blog/products/ai-machine-learning/an-in-depth-look-at-googles-first-tensor-processing-unit-tpu},
  2017.

\bibitem{Shutz18}
L.~Schulz, D.~Broneske, and G.~Saake.
\newblock An {E}ight-{D}imensional {S}ystematic {E}valuation of {O}ptimized
  {S}earch {A}lgorithms on {M}odern {P}rocessors.
\newblock {\em Proc. VLDB Endow.}, 11:1550–1562, 2018.

\bibitem{vaidya2020partitioned}
K.~Vaidya, E.~Knorr, T.~Kraska, and M.~Mitzenmacher.
\newblock Partitioned learned bloom filter.
\newblock {\em ArXiv}, abs/2006.03176, 2020.

\bibitem{LawsDead}
B.~Wang.
\newblock Moore's {L}aw is {D}ead but {GPU} {W}ill {G}et 1000x {F}aster {B}y
  2025.
\newblock
  \url{https://www.nextbigfuture.com/2017/06/moore-law-is-dead-but-gpu-will-get-1000x-faster-by-2025.html},
  2017.

\end{thebibliography}



\end{document}